\documentstyle[preprint,aps,floats,epsfig,subeqn,amssymb]{revtex}

\newcommand{\be}{\begin{equation}}
\newcommand{\ee}{\end{equation}}
\newcommand{\bea}{\begin{eqnarray}}
\newcommand{\eea}{\end{eqnarray}}
\newcommand{\bml}{\begin{mathletters}}
\newcommand{\eml}{\end{mathletters}}
\newcommand{\re}[1]{(\ref{#1})}

\begin{document}

\tighten

\preprint{DCPT-03/01}
\draft




\title{Higher order curvature generalisations of Bartnick-McKinnon and coloured
black hole solutions in $d=5$}
\renewcommand{\thefootnote}{\fnsymbol{footnote}}
\author{ Yves Brihaye\footnote{Yves.Brihaye@umh.ac.be}}
\address{Facult\'e des Sciences, Universit\'e de Mons-Hainaut,
7000 Mons, Belgium}
\author{A. Chakrabarti,\footnote{chakra@cpht.polytechnique.fr}}
\address{ Laboratoire de Physique
Th\'eorique, Ecole Polytechnique, Palaiseau, France}
\author{Betti Hartmann\footnote{Betti.Hartmann@durham.ac.uk}}
\address{Department of Mathematical Sciences, University
of Durham, Durham DH1 3LE, U.K.}
\author{D. H. Tchrakian\footnote{tigran@thphys.may.ie}}
\address{Department of
Mathematical Physics, National University of Ireland Maynooth, Ireland\\
and\\
School of Theoretical Physics -- DIAS, 10 Burlington
Road, Dublin 4, Ireland}
\date{\today}
\setlength{\footnotesep}{0.5\footnotesep}

\maketitle
\begin{abstract}
We construct globally regular as well as non-abelian black hole
solutions of a higher order curvature Einstein-Yang-Mills (EYM) model in
$d=5$ dimensions. This model
consists of the superposition of the first two members of the
gravitational hierarchy (Einstein plus first Gauss-Bonnet(GB)) interacting
with the superposition of the first two members of the $SO(d)$
Yang--Mills hierarchy. 
\end{abstract}

\pacs{PACS numbers: 04.50.+h, 04.40.Nr, 04.70.Bw, 11.10.Kk }

\renewcommand{\thefootnote}{\arabic{footnote}}
\section{Introduction}
In a previous paper~\cite{bct} we discussed regular solutions to the
$d$ dimensional Einstein--Yang-Mills (EYM) system consisting of the first
two members of each hierarchy in increasing orders of the curvatures, the
YM fields with gauge group $SO(d)$. Concretely, we constructed these
solutions for $d=6,7$ and $8$. In \cite{bct} we restricted to those
dimensions in which the solutions constructed had nontrivial
flat-space limits. Due to Derrick's scaling requirement, $d=5$ was
excluded. Here we consider this system in dimension
$d=5$, and in addition to regular ones, we construct also
black hole solutions.

It is well known that Derrick's theorem forbids static, globally regular,
finite energy solutions in pure YM theory, and that 
pure Einstein gravity has also no globally regular solutions. Finite
energy solutions in the combined EYM system however, are not obstructed
by this scaling requirement and indeed spherically symmetric
particle-like solutions in this system
in $d=4$ dimensions \cite{BM} were discovered. In fact, Bartnik and
McKinnon discovered an infinite series of solutions indexed by the number
$k$ of nodes of the gauge field function. The energy of these solutions
is however not bounded from below by any topological charge, and as
expected are in this sense sphaleron- like \cite{gv1}.
Shortly after this discovery, the corresponding black hole solutions,
so-called ``coloured black holes'' where constructed \cite{vg,km,bizon}.
These provided one of the first examples of the fact that the famous
``No-hair'' conjecture cannot be generalised to 
theories which involve non-linear matter sources. In the years that
followed, solutions of $d=4$ EYM theory have been studied in great
detail (see \cite{gv} for a review).

Field theories in more than $4$ dimensions gained increasing interest
in past years since it now seems accepted that many aspects of field
theory depend on the higher (than four) dimensional aspects of some
of these theories.  
Superstring theory \cite{string}, which is considered to be a good
candidate for a ``Theory of Everything'' (TOE),
is only consistent in $d=10$. In the low energy effective action limit
of (Super)string theories, new terms appear. In the gravity sector,
these are the Gauss-Bonnet terms, while e.g.
for Yang-Mills fields a hierarchy appears, which results from the
replacement of the standard Yang-Mills action by the corresponding
Born-Infeld action.

Recently moreover, the Randall-Sundrum (RS) models \cite{rs1,rs2} 
have pushed forward the study of solutions
in dimensions $d >4$ even further. These models involve either
two (RSI) \cite{rs1} or one (RSII) \cite{rs2} 3-brane(s) in a
5-dimensional space-time. While gravity ``lives'' in the full 5
dimensions, the matter fields are confined to the brane. 

Consequently, Einstein-Yang-Mills solutions in $d=5$ have been studied
\cite{volkov,maeda}. In \cite{volkov} it was found that no static,
globally regular, finite energy solutions exist  in an asymptotically
flat space-time. Only if $\partial/\partial x_4$ with $x_4$ being the
coordinate of the 5th dimension is chosen 
to be a Killing vector of the model, finite energy solutions are possible.
In \cite{maeda} globally regular as well as non-abelian black hole
solutions of the $d=5$ EYM model in asymptotically flat, in de Sitter (dS)
and in Anti-de Sitter (AdS) space-time have been constructed. It was
found that -like in asymptotically flat space-
non globally regular, finite energy solutions are possible
in  a space-time with negative cosmological constant (AdS space).

We give the model, ansatz, classical field equations and boundary
conditions for generic $d$ in Section II. We discuss
Reissner-Nordstr\"om-like solutions in Section III, the generalised
Bartnik-McKinnon solutions in Section IV
and the generalised coloured black hole solutions in Section V.
We give our conclusions in Section VI.

\section{The model}
We consider here the Lagrangian employed in \cite{bct}, containing only
the first two terms of both the gravitational and Yang-Mills hierarchies,
namely
\begin{equation}
{\cal L}={\cal L}_{grav}+{\cal L}_{YM}
\end{equation}
with
\begin{equation}
{\cal L}_{grav}=\sqrt{-g}\left(\frac{\kappa_1}{2}R_{(1)}+
\frac{\kappa_2}{4} R_{(2)}\right)
\end{equation}
and
\begin{equation}
{\cal L}_{YM}=\sqrt{-g}\left(\frac{\tau_1}{4}\mbox{T}r F^2(2)+
\frac{\tau_2}{48}\mbox{Tr} F^2(4)\right) \ .
\end{equation}
$g$ denotes the determinant of the $d$-dimensional metric tensor,
$R_{(1)}$ is the standard Ricci scalar appearing in the Einstein-Hilbert
action, while $R_{(2)}$ is the $2$- Ricci scalar associated with the
first Gauss-Bonnet (GB) term. $F(2)$ denotes the standard Yang-Mills
2-form, while $F(4)$ is the Yang-Mills 4-form resulting from
the total antisymmetrisation of $F(2)$ (see \cite{bct} for details).

Though we are primarily interested in $d=5$ in this paper,
we give the ansatz, the equations and the boundary conditions in this
section for generic $d$. This is because, before proceeding to
construct solutions numerically in Sections {\bf IV} and {\bf V} in $d=5$,
we will consider the question of Reissner-Nordstr\"om-like
solutions in section {\bf III}, for generic $d$.

\subsection{The Ansatz}

The choice of gauge group is somewhat flexible. In \cite{bct} the gauge
group chosen was $SO(d)$, in $d$ dimensions. But the gauge field of
the static solutions in question took their values in $SO(d-1)$. Thus in
effect, it is possible to choose $SO(d)$ in the first place. Now for {\it
even} $d$, it is convenient to choose $SO(d)$ since we can then avail
of the chiral representations of the latter, although this is by no means
obligatory. Adopting this criterion, namely to employ chiral
representations, also for {\it odd} $d$, it is convenient to choose the
gauge group to be $SO(d-1)$. We shall therefore denote our representation
matrices by $SO_{\pm}(\bar d)$, where $\bar d=d$ and $\bar d=d-1$ for
{\it even} and {\it odd} $d$ respectively.

In this unified notation (for odd and even $d$), the spherically symmetric
Ansatz for the $SO_{\pm}(\bar d)$-valued gauge fields then reads:
\begin{equation}
\label{YMsph}
A_0=0\ ,\quad
A_i=\left(\frac{w(r)-1}{r}\right)\Sigma_{ij}^{(\pm)}\hat x_j\ , \quad
\Sigma_{ij}^{(\pm)}=
-\frac{1}{4}\left(\frac{1\pm\Gamma_{ \bar d+1}}{2}\right)
[\Gamma_i ,\Gamma_j]\ .
\end{equation}
The $\Gamma$'s denote the $\bar d$-dimensional gamma matrices and
$1,j=1,2,...,d-1$ for both cases.

For the metric, we choose the Schwarzschild-type spherically symmetric
Ansatz:
\begin{equation}
ds^2=-\sigma^2(r)N(r)dt^2+N^{-1}(r)dr^2+r^2 d\Omega_{d-2}^2
\end{equation}
where $\sigma$ and $N$ are only functions of the $d-2$ dimensional radial
coordinate $r$. The consistency of this Ansatz for this system has been
checked~\cite{bct}.The function $N(r)$ is related to the mass function
$m(r)$ in the following way:
\begin{equation}
\label{mdef}
m(r)=n_d^{-1}[\kappa_1 r^{d-3}(1-N)+\frac14\kappa_2 r^{d-5}(1-N)^2]\ ,
\end{equation}
where $n_d$ is the dimension of the $\Gamma$ matrices entering in (\ref{YMsph}).
$n_d=2^{(d-2)/2}$ for $d$ even and $n_d=2^{(d-1)/2}$ for $d$ odd.
For $\kappa_2=0$ and $d=4$ (\ref{mdef}) reduces to the 
standard definition of the mass function $m(r)$.

\subsection{The classical field equations}
Inserting the Ansatz into the action and varying with respect to the
matter field $w(r)$ and the metric fields $N(r)$, $\sigma(r)$, we obtain
the classical field equations~: 
\begin{eqnarray}
&2\tau_1&\left(\left(r^{d-4}\sigma Nw'\right)'
-(d-3)r^{d-6}\sigma(w^2-1)w\right)+\nonumber \\
+&3\tau_2&(d-3)(d-4)(w^2-1)\left(
\left(r^{d-8}\sigma N(w^2-1)w'\right)'
-(d-5)r^{d-10}\sigma(w^2-1)^2w\right)=0
\label{YM12eq}
\end{eqnarray}
\begin{eqnarray}
m'&=&\frac18r^{d-4}\Bigg(\tau_1\left[Nw'^2
+\frac12(d-3)\left(\frac{w^2-1}{r}\right)^2\right] \nonumber \\
&+&\frac32\frac{\tau_2}{r^2}(d-3)(d-4)\left(\frac{w^2-1}{r}\right)^2
\left[Nw'^2+\frac14(d-5)\left(\frac{w^2-1}{r}\right)^2\right]\Bigg)
\label{meq}
\end{eqnarray}
\begin{equation}
\left[\kappa_1+\frac{\kappa_2}{2r^2}(d-3)(d-4)(1-N)\right]\left(
\frac{\sigma'}{\sigma}\right)=\frac{n_d}{8r}
\left[\tau_1+\frac32\frac{\tau_2}{r^2}(d-3)(d-4)
\left(\frac{w^2-1}{r}\right)^2\right]w'^2\ .
\label{sigeq}
\end{equation}
For $\kappa_2=\tau_2=0$ and
$d=4$, (\ref{YM12eq})-(\ref{sigeq}) coincide with the ordinary
differential equations of \cite{BM,gv}.

\section{Reissner-Nordstr\"om-like solutions}
For $d=4$ the Reissner-Nordstr\"om (RN) solution is a solution
of the Einstein-Yang-Mills equations \cite{RNEYM} with
$w\equiv 0$ and $\sigma\equiv 1$.
Interestingly, it was found that the RN solution is the limiting solution
of the sequence of Bartnik-McKinnon (BM) solutions for the number of
gauge field nodes going to infinity. Thus, a charged solution (the RN
solution) represents the limiting
solution of a sequence of non-charged  solutions (the BM solutions).
However, it should be noted that through suitable rescalings in the
$d=4$ EYM system, the only parameter that can be changed for spherically
symmetric solutions is the number of nodes of the gauge field. In this
paper, we have additional parameters, which result from the inclusion
of higher order curvature and
Yang-Mills hierarchy terms. Here we study whether the RN solution
is still a possible solution of the generalised EYM equations. 

In this Section, we give an answer to this question, for generic $d$.
For our considerations, it is immaterial how many GB terms the
gravitational sector contains. What is important is the YM content of
increasing orders. For this purpose, it is useful to consider the
most general version of equation \re{meq}, but in practice we restrict
to the case with the first three members of the YM hierarchy, from which
it is easy to conclude the maximal superposition of YM terms\footnote{By
maximal superposition we mean the superposition of $F(2p)^2$ terms for
up to $p=\frac{d}{2}$ for even $d$ and $p=\frac{d-1}{2}$ for odd
$d$, above which these terms become total divergences.}.

The generalisation of \re{meq} for the case with YM content up to the
$p=3$ member is
\bea
m'&=&\frac18r^{d-2}\Bigg(\tau_1\left[N\left(\frac{w'}{r}\right)^2
+\frac12(d-3)\left(\frac{w^2-1}{r^2}\right)^2\right] \nonumber \\
&+&\frac32\tau_2(d-3)(d-4)\left(\frac{w^2-1}{r^2}\right)^2
\left[N\left(\frac{w'}{r}\right)^2+\frac14(d-5)
\left(\frac{w^2-1}{r^2}\right)^2\right]\nonumber\\
&+&\frac{147}{160}\tau_3(d-3)(d-4)(d-5)(d-6)
\left(\frac{w^2-1}{r^2}\right)^4
\left[N\left(\frac{w'}{r}\right)^2+\frac16(d-7)
\left(\frac{w^2-1}{r^2}\right)^2\right]\Bigg)
\label{meqgen}
\eea
in which the mass function $m(r)$ can incorporate all possible GB terms
allowed in the given dimensions~\cite{ct}.

Setting $w(r)\equiv 0$ in \re{meqgen}, and then integrating it,
we find for $d=5$:
\begin{equation}
m(r)=\frac{\tau_1}{8} \ln r + C_5
\end{equation}
while for $d=9$, we have:
\begin{equation}
m(r)=\frac{3}{32} \tau_1 r^4 +\frac{45}{8}\tau_2 \ln r -\frac{441}{128} \tau_3 r^{-4} + C_9
\end{equation}
and for $d=13$:
\begin{equation}
m(r)=\frac{5}{16} \tau_1 r^8 + \frac{135}{4} r^4 + \frac{9261}{16} \ln r + C_{13}
\end{equation}
with $C_5$, $C_9$ and $C_{13}$ being the respective integration constants.
Clearly, for $d=5$, $9$, $13$, these solutions have infinite energy and are not regular
at the origin. A $d=5$ solution with logarithmically divergent mass was also found in
\cite{maeda} and was named ``quasi-asymptotically'' flat.  

We can look for horizons of these solutions, at which $N(r=r_h)=0$. For $d=5$ and $\kappa_2=0$ (which is the case,
we are interested in mainly in this paper), we then have from 
(\ref{mdef}) 
\begin{equation}
N(r_h=0) \ \ \Rightarrow \ \ \frac{n_d}{\kappa_1 r^2}\left(\frac{\tau_1}{8}\ln(r)+C_5\right)=1
\end{equation}
which has solutions if $C_5 > 0$. For an extremal solution, for which in addition we have $\frac{\partial N}{\partial r}\vert_{r=r^{ex}_h}=0$, we find
that
\begin{equation}
r_h^{ex}=\sqrt{\frac{n_d \tau_1}{16\kappa_1}} \ .
\end{equation}
Thus, indeed, magnetically charged black hole solutions exist in $d=5$. However, these
have logarithmically divergent energy.

For $d\neq 5$, $9$, $13$ we find from \re{meqgen}:
\bea
r^{-d+1}m(r)&=&\frac{d-3}{16}\Bigg(Cr^{-d+1}+\frac{\tau_1}{d-5}r^{-4}
+\frac34\ \frac{\tau_2}{d-9}(d-4)(d-5)r^{-8}\nonumber\\
&+&\frac{49}{160}\ \frac{\tau_3}{d-13}(d-4)(d-5)(d-6)(d-7)r^{-12}\Bigg)
\label{RNcondition}
\eea
where $C$ is an integration constant.

Now gravitational equations,
for example \re{meqgen}, support RN-type solutions~\cite{MP,ct}
if the right hand side of \re{RNcondition} is proportional to :
\begin{equation}
\label{powers}
a r^{-d+1} - b r^{4-2d} \ , \ \ a,b \ \ \rm{constants}  \ .
\end{equation}
The $r^{-d+1}$ expression is the Schwarzschild-like term and $a$ is
related to the mass of the solution, while the $r^{4-2d}$ part denotes
a RN type term with $b$ being related to the global charge.
Comparing the powers of $r$ on the right hand side of \re{RNcondition} with
\re{powers} shows that in the example at hand there are only three cases
wherethis can be fulfilled for $a\neq 0$, $b\neq 0$
\begin{itemize}
\item $\tau_2=\tau_3=0$ and $d=4$: This is the chase of the usual EYM
equations in $4$ dimensions \cite{BM}, which was shown to have an
embedded RN solution \cite{RNEYM}.
\item $\tau_1=\tau_3=0$ and $d=6$: In this case, an embedded RN solution
is possible, however due to the
absence of the standard YM term it seems rather unphysical. 

\item $\tau_1=\tau_2=0$ and $d=8$: In this case too, an embedded RN
solution is possible, which again is rather unphysical in the
absence of the standard YM term.
\end{itemize} 

It follows by induction from the above argument that RN type solutions
can be embedded by the above prescription, {\it only in those} generalised
EYM {\it systems consisting of a single} YM {\it member, namely that with
$p=\frac{d-2}{2}$.} One conclusion that may follow from this is, that
for EYM systems with multi-member YM sector, there may be no multi-node
solutions at all, since in $d=4$ the existence of multinode solutions is
initmately tied up with the existence of limiting RN field
configurations.

For all other cases, no RN type solutions are possible. However, there
exists an embedded solution for all $d$, namely the Schwarzschild
solution. This can be easily seen by setting $w=\pm 1$ in (\ref{meq})
and using the relation between $m$ and $N$.

\section{Generalised Bartnik-McKinnon solutions}

\subsection{The boundary conditions}
We require asymptotic flatness
and finite energy. Thus the boundary conditions at infinity read:
\begin{equation}
\label{bcinf}
\sigma(r=\infty)=1 \ , \  \  \ w(r=\infty)=(-1)^{k} 
\end{equation}
with $k=1,2,3,....$ being the node number of the gauge field function $w(r)$.
For odd $k$, the gauge field function at infinity behaves like:
\begin{equation}
 w(r\rightarrow \infty) = -1 + \frac{C}{r^{d-3}} \ , 
\end{equation}
while for even $k$, we have:
\begin{equation}
 w(r\rightarrow \infty) = 1 - \frac{C}{r^{d-3}}  \ .
\end{equation}

The requirement of regularity at the origin leads to the
conditions \cite{bct}~:
\begin{equation}
m(0)=0 \ , \ \ \ w(0)=1 \ .
\end{equation}
From here on, we specialise to the spacetime dimension $d=5$.

\subsection{Numerical results}
In this section we present the results we obtained for the two
cases corresponding to $p=1$ and $p=2$ gravity in $d=5$ space-time
dimensions.

When the two mixing parameters $\tau_1, \tau_2$ are non-zero 
we can set them equal to particular values without losing
generality. This can be done by an appropriate
rescaling of the overall Lagrangian density and of the radial variable.
So we take advantage of this freedom
and choose $\tau_1 = 8, \tau_2 = 8/3$ (this simplifies the numerical
coefficients in the equations).

\subsubsection{$\kappa_1\neq 0$, $\kappa_2=0$}
This case corresponds to the case of pure Einstein-Hilbert gravity.
We define $\alpha^2  \equiv n_d / (2 \kappa_1)$. The limit
$\alpha^2 \rightarrow 0$ then corresponds to the flat case.
As was already noticed in \cite{bct}, Derrick's 
theorem does not allow regular solutions
in the flat limit, and our numerical analysis indeed confirms this
expectation. However for $\alpha^2 > 0$,  we were able to construct
regular solutions. This is remarkable, since it was observed in
\cite{volkov,maeda}, that no finite energy solutions exist for $\tau_2=0$.
As expected the inclusion of the second term of the Yang-Mills hierarchy
leads to the possibility of having finite energy solutions.
These solutions exist up to a maximal
value, $\alpha_{max}$, of the parameter $\alpha$. We find numerically
$\alpha_{max}^2 \approx 0.28242$.

Although no regular solution exist in the flat limit, 
some comments are in order here.  In our equations only the 
coupling to gravity fixes the scale and prevents the solution
from shrinking to zero.  Rescaling the radial variable
according to  $y = \alpha r$, we see that the right hand side
of (\ref{sigeq}) vanishes in the $\alpha \rightarrow 0$
limit while the term proportionnal to $\tau_2$ in (\ref{YM12eq}) gets
multiplied by $\alpha^4$. Therefore, taking the limit $\alpha =0$
leads to  the following form for (\ref{YM12eq})~:
\begin{equation}
\label{instanton}
\frac{d}{dy} (y \frac{d w}{dy}) - 2 w (w^2-1) = 0
\end{equation}
which is nothing else but the equation of the YM
instanton~\cite{instanton} in $4$-dimensional flat space. Its solution
is $w(y) = (1-y^2)/(1+y^2)$ and it has
mass $M= 8/3$, in complete agreement with our numerical solution.

When $\alpha^2$ increases, we observe that the mass decreases, and that
both, the value $\sigma(0)$ and the minimum $N_m$ of the function $N(r)$
decrease, as indicated in Fig.~1. 

The stopping of the main branch of solutions at
$\alpha^2 = \alpha_{max}^2$
strongly suggest the existence of a second branch of solutions. We
indeed confirmed this numerically. We found another branch of solutions
on the interval $\alpha^2 \in [\alpha_{cr(1)}^2 , \alpha_{max}^2]$ 
with $\alpha_{cr(1)}^2 \approx 0.1749$.
On this second branch of solutions, both $\sigma(0)$ and $N_m$ continue
to decrease but stay finite. However, a third branch of solutions exists
for $\alpha^2 \in [\alpha_{cr(1)}^2 , \alpha_{cr(2)}^2]$ , 
$\alpha_{cr(2)}^2 \approx 0.1778$ on which the two quantities decrease
further. Progressing on this succession of branches the main observation
is that the value $\sigma(0)$ decreases much quicker that $N_m$ 
as illustrated in Fig.~1.  In Fig.~2, we compare the profiles of
the functions $N$, $\sigma$ and $w$ for the same value of
$\alpha^2$ on the first and third branch.  The pattern
strongly suggests that after a finite (or infinite) number of branches
the solution terminates into a singular solution with $\sigma(0)=0$.
The existence of a series of branches in $d=5$, albeit in a different
model, has already been observed in \cite{volkov}.

Since our equations generalise the Bartnik-McKinnon (BM) equations, 
which admit a sequence of
regular, finite energy solutions
indexed by the number of nodes of the gauge function $w(r)$
, we have tried to construct the counterparts of these 
multi-node solutions for $d=5$.  However, we have not succeeded in
constructing them.  
Our numerical technique was to start from the two-node BM solution
for $d=4$ and to solve the equations while increasing the dimension
$d$ gradually. At $d=4$ the $p=2$ YM term trivialises, but when $d$
increases to values higher than $4$, the $p=2$ YM terms comes into
play. We found that the BM solution is indeed
deformed for $5 > d>4$ and that the position of the two nodes
is a function of $d$. In the limit $d \rightarrow 5$
our numerical procedure indicates that the position of the second node
reaches out to infinity. This provides an argument that - if multi-node
solution would exist at all in the $d=5$ case - it will be very difficult
to construct them.

\subsubsection{$\kappa_1=0$, $\kappa_2\neq 0$}
When gravity just reduces to the Gauss-Bonnet term
($\kappa_1 =0$) then from (\ref{mdef}) 
we find  the relation between $N$ and $m$ to be:
\begin{equation}
1-2\sqrt{m(r)\frac{4}{\kappa_2}}=N(r) \ .
\end{equation}
Clearly, this is {\bf not}  an asymptotically flat solution, since $m(\infty)$,
which is proportional to the mass of the solution, should be non-zero.
For large $r$, we can assume the mass function to be approximately
constant $m(r >> 1)\approx m_{\infty}=const.$. Moreover, 
we can assume $\sigma(r >> 1)\approx 1$.
Following a similar argument as in \cite{vb} and rescaling the $t$ and 
$r$ coordinate, we end up with a metric which has a deficit angle:
\begin{equation}
ds^2=-d\tilde{t}^2 +  d\tilde{r}^2 + \left(1-2
\sqrt{m_{\infty}\frac{4}{\kappa_2}}\right) \tilde{r}^2 d\Omega^2_{3} \ .
\end{equation}
Thus a $3$ dimensional hypersphere of radius $r$ has surface
area $2\pi^2 r^3\left(1-2\sqrt{m_{\infty}\frac{4}{\kappa_2}}\right)$.
 
\subsubsection{$\kappa_1\neq 0$, $\kappa_2\neq 0$}
Considering small values of $\kappa_2$, the solutions
hardly differ from the pure Einstein-Hilbert gravity solutions.
When the Gauss-Bonnet coupling constant becomes of the same order
as $\kappa_1$  noticable differences appear.
For instance, for $\kappa_2 = \kappa_1$ we have constructed 
solutions on the first branch up to $\alpha^2 \approx 0.5$.
The  construction of further branches in this case appears to
be numerically very difficult.

\section{Generalised Coloured Black hole solutions}

\subsection{Boundary conditions}
Since we are still considering asymptotically flat, finite energy
solutions, the boundary conditions (\ref{bcinf}) are till valid.

Furthermore, black holes solutions are characterized by an horizon
$r\equiv x_h$ at which the metric function $N$ vanishes~: $N(x_h) = 0$.
This leads (for $\kappa_2=0$) to the condition~:
\begin{equation}
\label{bh1}
       m(x_h) = \frac{\kappa_1}{n_D} x_h^{d-3} \ .
\end{equation}

The requirement of regularity of the gauge function $w(r)$ at $x_h$
leads to the following condition~: 
\begin{eqnarray}
\label{bh2}
  && \tau_1 (r^6 N' w' - (d-3) r^4 w (w^2-1)) \nonumber  \\
&+&  \frac{\tau_2}{2}(d-4)(d-3) (w^2-1) 
(r^2 N' w' (w^2-1) - (d-5) w(w^2-1)^2)\vert_{r = x_h} = 0 \ .
\end{eqnarray}

\subsection{Horizon properties}
The surface gravity \cite{sg} of the $d=5$ black hole, which
is proportional to the temperature $T$ of the black hole: $T=\kappa_{sg}/(2\pi)$,
is given by:
\begin{equation}
\kappa_{sg}^2=-\frac{1}{4}g^{tt} g^{rr} (\partial_r g_{tt})^2
\end{equation}
which gives:
\begin{equation}
\label{sg}
\kappa_{sg}=\sigma_{x_h}\left(\frac{1}{x_h}-\frac{\alpha^2}{x_h^3}(w_{x_h}^2-1)^2\right) \ ,
\end{equation}
where $\sigma_{x_h}$ and $w_{x_h}$, respectively, denote the values of
the metric function $\sigma$ and the gauge field function $w$ at the horizon.
As expected for spherically symmetric solutions, the expression
(\ref{sg}) clearly shows that the surface gravity is constant at the horizon,
thus the non-abelian black hole solutions fulfill the 0. Law of black hole
mechanics \cite{sg}.

Like the Gaussian curvature given by $R$, the curvature of a black hole
can also be described by the Kretschmann scalar.
For the 5-dimensional Kretschmann scalar $K=R^{ABCD}R_{ABCD}$,$A, B, C, D
=0,1,2,3,4$, at the horizon $x_h$
we find:
\begin{equation}
\label{kretsch}
K|_{x_h}=9 (N^{'}|_{x_h})^2 (\frac{\sigma^{'}}{\sigma}|_{x_h})^2+
6N^{'}|_{x_h}N^{''}|_{x_h}\frac{\sigma^{'}}{\sigma}|_{x_h}+
(N^{''}|_{x_h})^2+6\frac{(N^{'}|_{x_h})^2}{x_h^2}+\frac{12}{x_h^4} \ .
\end{equation}
This can be further evaluated by inserting the following expressions
into (\ref{kretsch}):
\begin{equation}
\frac{\sigma^{'}}{\sigma}|_{x_h}=2\alpha^2 (w^{'}|_{x_h})^2\left( 1+ 
\frac{(w_{x_h}^2-1)^2}{x_h^4}\right) \ ,
\end{equation}
\begin{equation}
N^{'}|_{x_h}=-\frac{2\alpha^2}{x_h^3}(w_{x_h}^2-1)^2+\frac{2}{x_h}
\end{equation}
and
\begin{equation}
N^{''}|_{x_h}=\frac{10\alpha^2}{x_h^4}(w_{x_h}^2-1)^2-
\frac{8\alpha^2 w_{x_h}
w^{'}|_{x_h}}{x_h^3}-\frac{2\alpha^2}{x_h}N^{'}|_{x_h}
(w^{'}|_{x_h})^2\left(
1+\frac{8}{x_h^4}(w_{x_h}^2-1)^2\right)
\end{equation}
We can thus express the Kretschmann scalar at the horizon in terms
of $w_{x_h}$, $w^{'}|_{x_h}$ and the horizon $x_h$ itself.
\subsection{Numerical results}
Here we limit our analysis to the case $\kappa_2 = 0$. 
We find that in analogy to the non-abelian black holes
in $d=4$ Einstein-Yang-Mills-Higgs (EYMH) theory \cite{bfm}, the $d=5$
black hole solutions exist in a limited domain of the
$\alpha$-$x_h$-plane. For small values of the gravitational coupling
$\alpha^{(4)}$, the EYMH solutions exist up to
a maximal value of the horizon $x^{(4)}_h=x^{(4)}_{h(max)}$ and on 
a second branch of solutions
bifurcate with the branch of non-extremal Reissner-Nordstr\"om (RN)
solutions at $x^{(4)}_h=x^{(4)}_{h(cr)}$. Both, $x^{(4)}_{h(max)}$
and $x^{(4)}_{h(cr)}$ are increasing functions of $\alpha^{(4)}$. On
the two branches, the value of the gauge field
function at the horizon
decreases monotonically from 1 to 0 at $x^{(4)}_{h(cr)}$ such
that the limiting solution is identical to the RN solution on the
full interval $r^{(4)}\epsilon [x^{(4)}_{h(cr)}:\infty]$.
For fixed $x^{(4)}_h$ and large $\alpha^{(4)}$, 
the situation changes and the solutions bifurcate with
the branch of extremal RN solutions forming a degenerate horizon for
$\alpha^{(4)}=\alpha^{(4)}_{(cr)}=x_{h(RN)}$. 
Now, however, the limiting solution
can only be described by the extremal RN solution for 
$r^{(4)}\epsilon [x^{(4)}_{h(RN)}:\infty]$. On the interval
$r^{(4)}\epsilon [x^{(4)}_{h}:x^{(4)}_{h(RN)}]$, the solution
is non-trivial and non-singular. 
Since (as argued in a preceeding section of this paper), there are no RN
solutions in $d=5$, we find a qualitatively different domain of existence
for the $d=5$ black holes. For fixed $\alpha$, we find the following
maximal values of $x_{h(max)}$ up to which the non-abelian 
solutions exist:

\begin{center}
Table 1 \\
\medskip
\begin{tabular}{|l||l|l|l|l|l|l|l|}
\hline 
$\alpha^2$ & $0.025$ & $0.05$ & $0.10$ & $0.15$ & $0.20$ & $0.25$ &
$0.27$\\
\hline \hline
$x_{h(max)}$ & $0.470$ & $0.468$ & $0.431$ & $0.385$ & $0.322$ & $0.220$ 
& $0.141$\\
\hline
\end{tabular}
\end{center} 

This describes rather a semi-circle than the complicated pattern found in
\cite{bfm}. Extending backwards in $x_h$, we find a second branch of
solutions for $x_h < x_{h(max)}$. Progressing on this branch, the value
$\sigma(0)$ drastically
decreases as shown in Fig.~3. We believe that the second branch stops
at $x^{(1)}_{h(cr)}\approx 0.2329$. This is indicated by the slight
``curling'' of $w_{x_h}$. We believe that a third branch exists on which
the value $\sigma(0)$
continues to decrease further to zero. However, it is likely that
the extension of this branch in $x_h$ will be very small, which might
be the reason why we were unable to find it so far. That we are close to
the critical solution is also indicated by the surface gravity
$\kappa_{sg}$. It drops down very quickly on the second branch and 
likely tends to zero in the limit of the critical solution. We have also 
evaluated the Kretschmann scalar at the horizon. We find that a sharp
increase in the curvature at the horizon is seen close to the critical
solution. We find, for example, that on 
the second branch $K|_{x_h}\approx 3970$ for $x_h=0.235$,
while $K|_{x_h}\approx 23992$ for $x_h=0.23295\approx x_{h(cr)}$.
 
In Fig.~4, we show the functions $w$, $\sigma$ and $N$
for the same values of $\alpha^2$ and $x_h$ on the 
two different branches. Clearly, these are distinct solutions.

Finally, in Fig.~5 we present the terminating solution that we were able
to construct on 
the second branch for $x_h=0.23295$. It is clearly seen in comparison to
Fig.~4 that $\sigma$ has a very large rise in value close to the horizon
and then in a small interval of the coordinate $r$ rises nearly linearly
to its asymptotic value $1$.

\section{Conclusions and Summary} 
The idea that our space-time consists of more than four dimensions
is not new and has been long discussed in string theory.
It is by now an accepted fact that bosonic string theory is only
consistent in $d=26$
space-time dimensions, while fermionic string theory ``lives'' in $d=10$
space-time dimensions.
Extra dimensions have recently gained additional interest in the context
of the Randall-Sundrum models \cite{rs1,rs2}. These are basically
$5$-dimensional space-times in which gravity lives in the full $5$
dimensions, while the remaining fields are confined to 3-branes. 

The study of soliton solutions in classical field
theories coupled to gravity leads in general
to a rich pattern of phenomena. A proof that
the study of non-abelian gauge fields interacting with gravity
is indeed worthwhile was the discovery of particle-like
solutions, the Bartnik-McKinnon solutions,
in the coupled Einstein-Yang-Mills (EYM) system \cite{BM}.
It is also remarkable that non-abelian black holes,
i.e. black holes which violate the ``No-hair'' conjecture,
can be constructed in non-abelian gauge theories coupled to gravity. 
One of the most famous examples are the coloured black hole solutions
constructed in EYM theory.

All this gives motivation to study the counterparts of the 
Bartnik-McKinnon and coloured black hole solutions 
in non-abelian gauge field theories
coupled gravity in more than four dimensions. Since in low
energy effective actions higher order curvature terms (Gauss-Bonnet terms)
as well as higher order terms from the Yang-Mills hierarchy
(Born-Infeld terms) appear, these should be taken into account when
studying solutions in $d > 4$. Here we have studied a $5$ dimensional
model which consists of the first two terms of the 
gravitational and Yang-Mills 
hierarchies, respectively. We have solved the system of coupled
differential equations numerically.

While without higher order Yang-Mills curvature terms
no globally regular, finite
energy solutions are possible \cite{volkov}, we find that in our model
they exist. A number of branches exist
which terminate into a singular solution. These solutions, which are
the analogues of the BK solution~\cite{BM} and are hence expected like
the latter to be sphaleron like, in other respects differ from these.
Most importantly these solutions exhibit branches more akin to those
in EYMH theory~\cite{bfm}. This is not too surprising since like the
latter, our model features dimensionful constants.

In passing we remark that unlike the BK solutions which consist of an
infinite sequence with multiple nodes, it seems most unlikely that there
should be any multinode solutions in any EYM model featuring more than
one YM term, as in our case.

We have also constructed non-abelian black holes. We find two branches 
of solutions. We were unable to reach the limiting solution
in this case, however our numerical results indicate that
the black hole solutions bifurcate with a singular solution
which has zero surface gravity and a very huge, if not infinite curvature
at the horizon.\\
\\
\\

{\bf Acknowledgements}
This work forms part of Enterprise--Ireland projects IC/02/005 and
SC/00/020. BH was supported by an EPSRC grant.
YB gratefully acknowledges the Belgian F.N.R.S. for
financial support.\\
\\
{\bf Note added}
We gratefully acknowlegde the unknow referee for his
constructive comments, especially for pointing out the
relation of our solutions with the 4-dimensional instanton.


\newpage

\begin{figure}
\centering
\epsfysize=20cm
\mbox{\epsffile{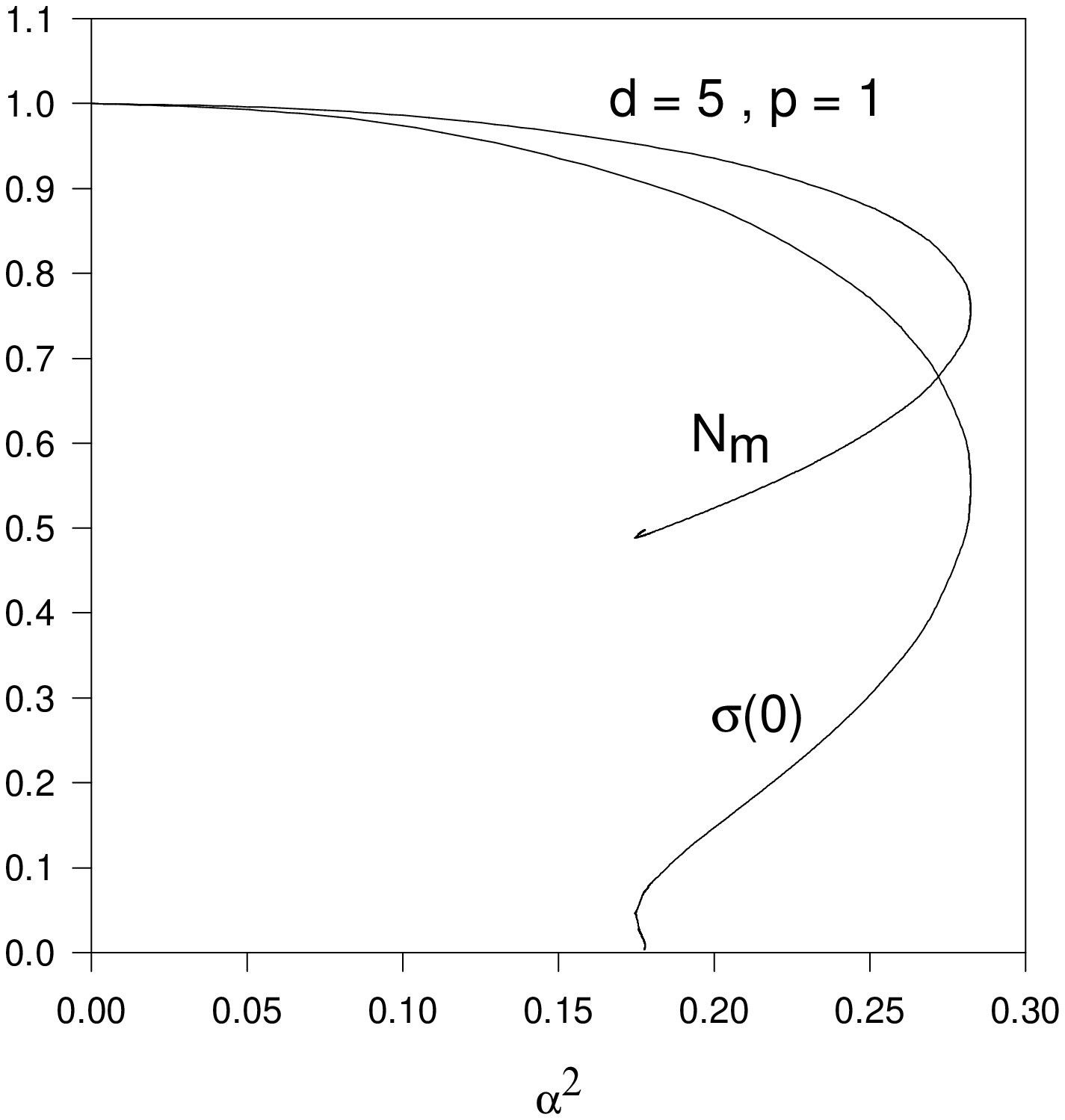}}
\caption{The value of the minimum of the metric function $N$, $N_m$, as
well as the value of the metric function $\sigma$ at the origin, $\sigma(0)$,
are shown for the generalised
Bartnik-McKinnon solutions
as functions of $\alpha^2=n_d/(8\kappa_1)$ for $p=1$ gravity ($\kappa_2=0$).
 }
\end{figure}
\newpage
\begin{figure}
\centering
\epsfysize=20cm
\mbox{\epsffile{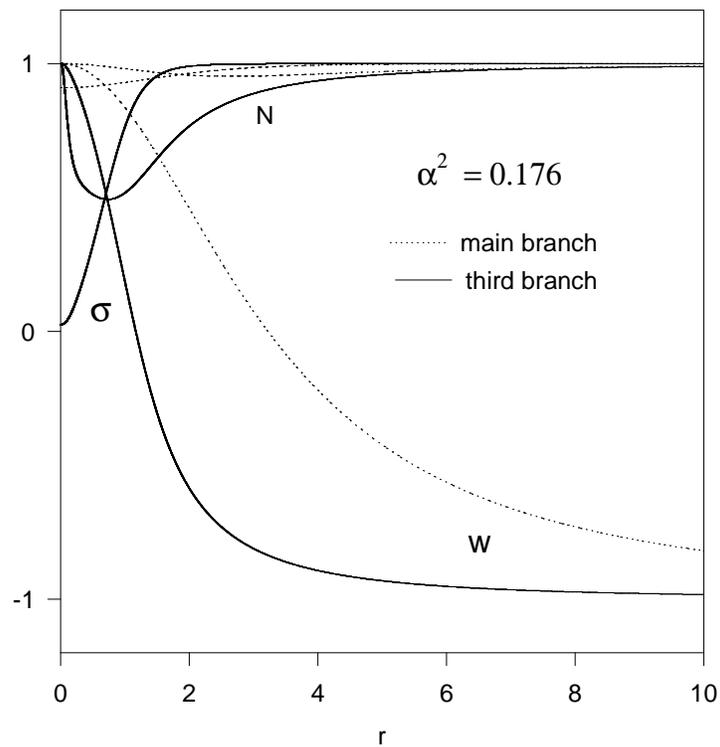}}
\caption{The gauge field function $w(r)$ as well as the metric functions
$N(r)$, $\sigma(r)$ are shown for the generalised
Bartnik-McKinnon solutions as functions of $r$ for $\alpha^2=0.176$ 
on the main and on the third branch of solutions, respectively.
 }
\end{figure}
\newpage
\begin{figure}
\centering
\epsfysize=20cm
\mbox{\epsffile{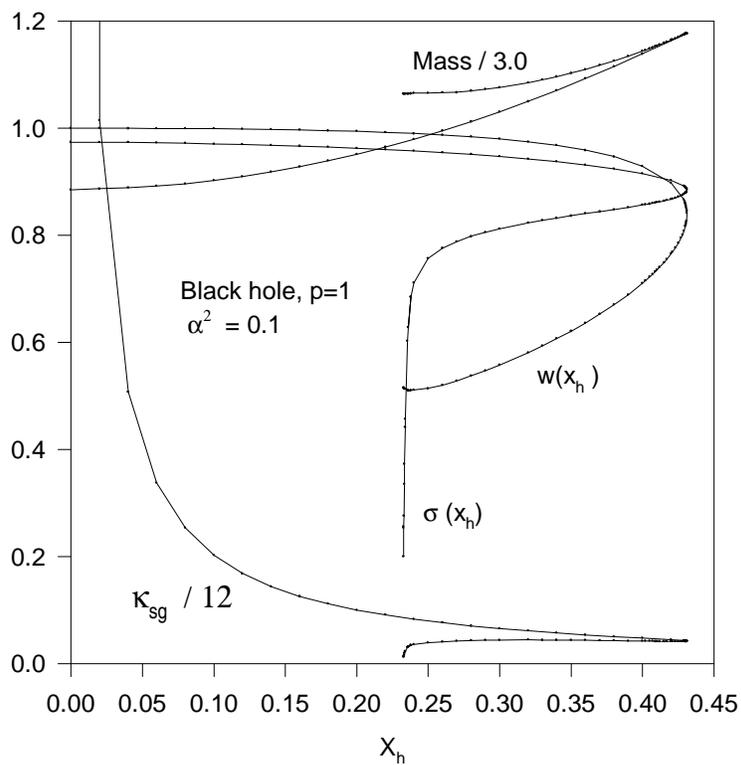}}
\caption{
The value of the gauge field function at the horizon $w(x_h)=w_{x_h}$,
the value of the metric function $\sigma$ at the horizon, 
$\sigma(x_h)=\sigma_{x_h}$, as well as the mass divided by 3 and the surface
gravity $\kappa_{sg}$ divided by 12 
are shown as functions of the horizon $x_h$ for
the generalised $d=5$ dimensional coloured black hole solutions with 
$\kappa_2=0$ and $\alpha^2=0.1$.
 }
\end{figure}
\newpage
\begin{figure}
\centering
\epsfysize=20cm
\mbox{\epsffile{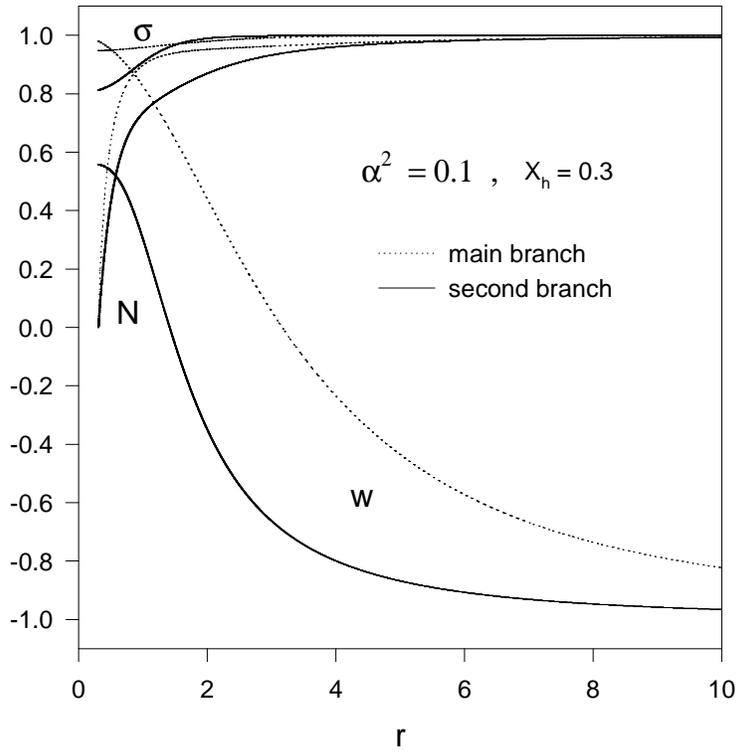}}
\caption{The gauge field function $w(r)$ as well as the metric functions
$N(r)$, $\sigma(r)$ are shown for the generalised
coloured black hole solutions as functions of $r$ for $\alpha^2=0.1$ 
and $x_h=0.3$ 
on the main and on the second branch of solutions, respectively.
 }
\end{figure}

\newpage
\begin{figure}
\centering
\epsfysize=20cm
\mbox{\epsffile{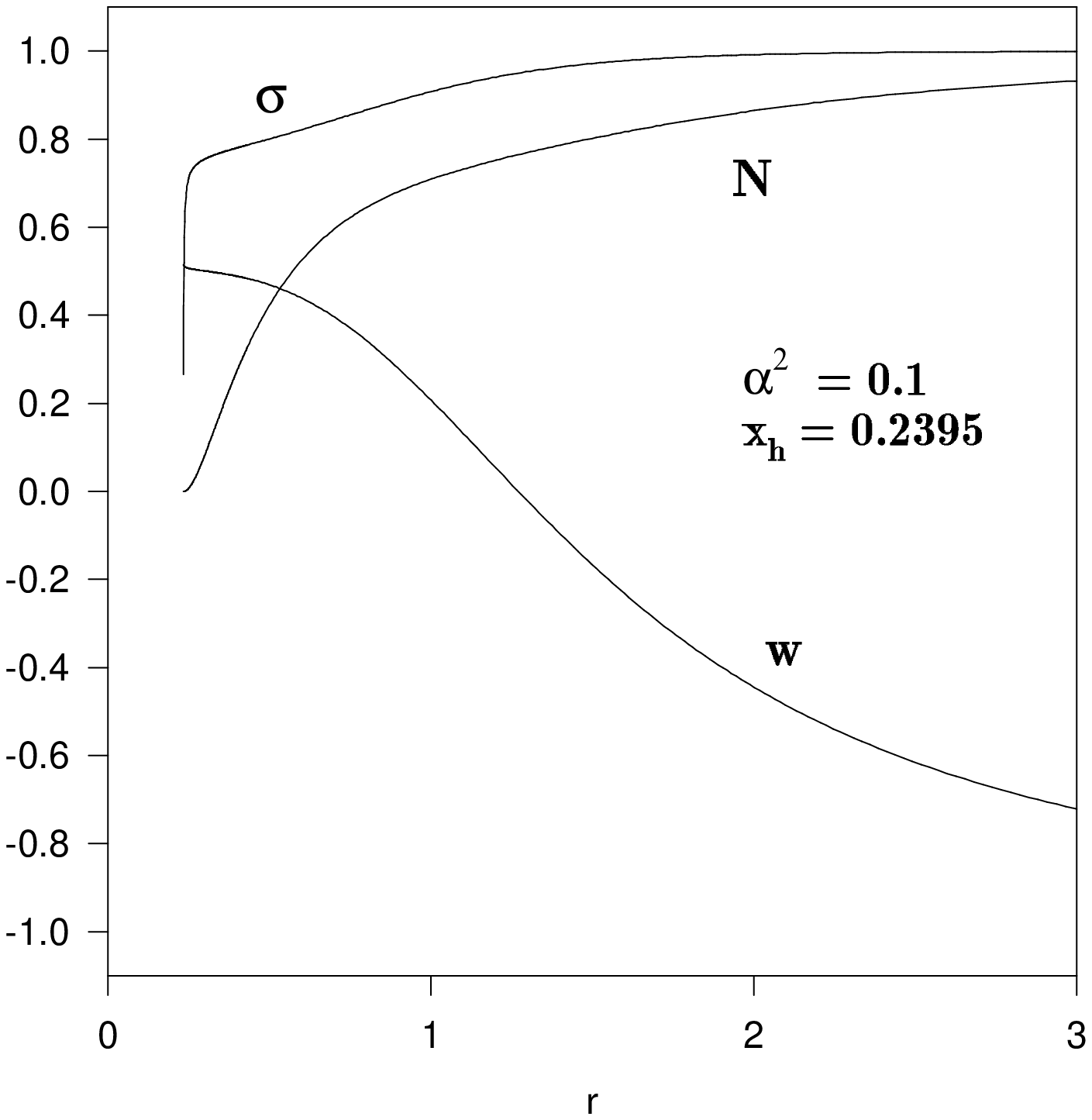}}
\caption{The gauge field function $w(r)$ as well as the metric functions
$N(r)$, $\sigma(r)$ are shown for the generalised
coloured black hole solutions as functions of $r$ for $\alpha^2=0.1$ 
and $x_h=0.23295$ close to the critical $x_h$.
 }
\end{figure}


\begin{thebibliography}{99}
\bibitem{bct} Y. Brihaye, A. Chakrabarti and D. H. Tchrakian,
Particle-like solutions to higher order curvature
Einstein--Yang-Mills systems in $d$ dimensions,
hep-th/0202141.
\bibitem{BM} R. Bartnik and J. McKinnon, Phys. Rev. Lett. {\bf 61}
(1988), 141.
\bibitem{gv1} D. Gal'tsov and M. Volkov, Phys. Lett. {\bf B 273} (1991),
255.
\bibitem{vg} M. Volkov and D. Gal'tsov, JETP Lett. {\bf 50} (1998), 346.
\bibitem{km} H. K\"unzle and A. Masood-ul-Alam, J. Math. Phys. {\bf 31}
(1990), 928.
\bibitem{bizon} P. Bizon, Phys. Rev. Lett. {\bf 64} (1990), 2844.
\bibitem{gv} M. Volkov and D. Gal'tsov, Phys. Rep. {\bf 319} (1999), 1.
\bibitem{string} see e.g. M. Green, J. Schwarz and E. Witten, 
{\it Superstring theory} (Vols. 1 and 2), Cambridge University Press
(1986); J.  Polchinski, {\it String Theory} (Vols. 1 and 2), Cambridge
University Press (1998).
\bibitem{rs1} L. Randall and R. Sundrum, Phys. Rev. Lett. {\bf 83}
(1999), 3370. 
\bibitem{rs2} L. Randall and R. Sundrum, Phys. Rev. Lett. {\bf 83}
(1999), 4690.
\bibitem{volkov} M. Volkov, Phys. Lett. {\bf B524} (2002), 369. 
\bibitem{maeda} N. Okuyama and K. Maeda,  Five-dimensional Black Hole and
Particle Solution with Non-Abelian Gauge Field, gr-qc/0212022.
\bibitem{ct} A. Chakrabarti and D. H. Tchrakian, Phys. Rev. {\bf D65}
(2002), 024029.
\bibitem{MP}
R.C. Myers and M.J. Perry, Ann. Phys. {\bf 172} (1986), 304.
\bibitem{RNEYM} J. A. Smoller and A. G.  Wasserman, J. Math. Phys.{\bf 38}
(1997), 6522.
\bibitem{vb} M. Bariolla and A. Vilenkin, Phys. Rev. Lett. {\bf 63}
(1989), 341.  
\bibitem{instanton}
A.A. Belavin, A.M. Polyakov, Yu.S. Tyupkin and A.S. Schwarz, Phys. Lett.
{\bf B 59} (1975) 85.
\bibitem{sg} see e.g. R. Wald, {\it General Relativity}, 
University of Chicago Press, Chicago, 1984.

\bibitem{bfm} P. Breitenlohner, P. Forgacs and D. Maison, Nucl. Phys.
{\bf B383} (1992), 357 ;
Nucl. Phys. {\bf B442} (1995), 126.
\end{thebibliography}
\end{document}